\documentclass[conference]{IEEEtran}

\usepackage{cite}
\usepackage{amsmath,amssymb,amsfonts}
\usepackage{algorithmic}
\usepackage{graphicx}
\usepackage{textcomp}
\usepackage{xcolor}
\usepackage{multirow}
\usepackage{subfig}

\definecolor{blue}{rgb}{0,0,0}
\definecolor{red}{rgb}{0,0,0}

\usepackage{url}
\def\BibTeX{{\rm B\kern-.05em{\sc i\kern-.025em b}\kern-.08em
    T\kern-.1667em\lower.7ex\hbox{E}\kern-.125emX}}

\begin{document}

\title{\textcolor{blue}{Adversarial Sample Generation for Anomaly Detection in Industrial Control Systems}

}


 \author{

     \IEEEauthorblockN{Abdul Mustafa\IEEEauthorrefmark{1}, Muhammad Talha Khan \IEEEauthorrefmark{1}, Muhammad Azmi Umer \IEEEauthorrefmark{2}, Zaki Masood \IEEEauthorrefmark{2}, Chuadhry Mujeeb Ahmed 
     \IEEEauthorrefmark{3}}
     \IEEEauthorblockA{\IEEEauthorrefmark{1}DHA Suffa University, Karachi, Pakistan
     \\\{cs172049, cs162021\}@dsu.edu.pk}
     \IEEEauthorblockA{\IEEEauthorrefmark{2}Singapore University of Technology and Design, Singapore
     \\\{azmi\_umer, zaki\_masood\}@sutd.edu.sg}
     \IEEEauthorblockA{\IEEEauthorrefmark{3}Newcastle University,  Newcastle, United Kingdom
     \\mujeeb.ahmed@newcastle.ac.uk}
     }


\maketitle


\begin{abstract}
Machine learning (ML)-based intrusion detection systems (IDS) are vulnerable to adversarial attacks. It is crucial for an IDS to learn to recognize adversarial examples before malicious entities exploit them. In this paper, we generated adversarial samples using the Jacobian Saliency Map Attack (JSMA). We validate the generalization and scalability of the adversarial samples to tackle a broad range of real attacks on Industrial Control Systems (ICS). We evaluated the impact by assessing multiple attacks generated using the proposed method. The model trained with adversarial samples detected attacks with 95\% accuracy on real-world attack data not used during training. The study was conducted using an operational secure water treatment (SWaT) testbed.
\end{abstract}


\section{Introduction}

Industrial control systems (ICS) comprise a \textcolor{red}{significant portion of any state or nation's critical infrastructure (CI)}. Examples of such systems include water treatment plants and electric power grids, where an ICS regulates the physical processes.  The physical processes consist of two primary parts: monitoring and controlling. The monitoring part maintains processes and ensures they are operating properly by measuring various signals acquired from sensors. The controlling part handles processes and makes decisions that enable actuators to perform actions \cite{perales2020madics}.

ICS and their modules were previously thought to be safe against cyber-attacks since they ran on proprietary hardware, software, and \textcolor{red}{air-gapped} networks that were not connected to the internet \cite{kravchik2018detecting}. However, as connectivity with the internet provides online access and monitoring functionalities, it has led to the necessity of connecting ICS components to other networks, subsequently contributing to the digitalization of industrial systems \cite{kravchik2018detecting}. 

Given their applicability, these systems have become a tempting target for attackers. Since these systems regulate real-world processes, cyber-attacks on them could have serious consequences for the ecosystems in which they are used, as well as for end-users \cite{ashibani2017cyber}.  As a result, it is evident that the security concerns of such systems are a serious global issue, and it is necessary to design robust systems to defend against cyber-attacks. Various security methods have been proposed for traditional IT systems, but applying them to ICS systems is complex since ICS devices have limited resources. They contain outdated systems and devices that do not support advanced safety mechanisms. Alternatively, security solutions such as passive monitoring of process data appear to be promising \cite{erba2019real}.

As a corollary, there has been a significant rise in research into ICS-tailored intrusion detection systems (IDS). These IDS monitor network or sensor data for attacks and anomalies that could impact ICS. Machine learning has seen a remarkable increase in use and integration with IDS due to its accuracy in detecting attacks~\cite{umer2022machine}. Unfortunately, the emergence of such systems has opened up a new attack vector, i.e., trained models can be targeted as well. Adversarial Machine Learning (AML) involves launching attacks on machine learning-based IDS models by exploiting flaws in the trained models, such as "blind spots" among training examples. Specifically, by introducing minor perturbations to data points not seen during training, it is possible to cross decision boundaries and classify data into different classes. As a consequence, the model's performance may suffer when encountering previously unseen data points, leading to an increase in misclassifications. AML can be used to manipulate data received via actuators and other devices. In the context of ICS, this is done by introducing perturbations that cause attack data to be categorized as normal, thereby evading the IDS. This could result in late detection of an attack, information leakage, financial loss, etc.

In the current study, we assess the effectiveness of adversarial samples generated using machine learning against cyber-attacks on a water treatment plant \cite{swat2016}. This work uses the \textcolor{blue}{secure water treatment (SWaT)} (December 2015) attacked dataset \cite{swatDataset} to generate adversarial samples using Jacobian saliency map attack (JSMA). \textcolor{red}{By utilizing JSMA, we can more accurately assess the effectiveness of the adversarial samples.} Although the adversarial samples are generated using attack data, however, this attack data was never used directly in training the machine learning classifiers. Therefore, the result of such a trained model with high accuracy is an indication of the usefulness of adversarial sample generation techniques, in attack detection.
\textcolor{red}{In this paper, we utilize a real-world SWaT attack dataset to provide a more realistic evaluation of our solution’s performance under different machine learning classifiers with varying accuracy.}

\textcolor{blue}{The main contributions of our paper are as follows:}
\begin{itemize}
    \item This paper presents an approach to generate synthetic adversarial samples using JSMA. 
    It also highlights that the JSMA which was originally designed for image media, could be extended to generate adversarial samples for time series data. 
    In particular, our approach is significant in understanding the potential weaknesses in current ML algorithms, particularly in security-sensitive applications.
    \item To validate our approach, we conducted various experiments demonstrating the practical effectiveness of synthetic adversarial samples against ML-based IDS. 
    Our results show a significant decrease in the detection rate of the IDS when exposed to these adversarial samples, underscoring the critical need for enhanced security measures in ML-based approaches.
\end{itemize}

\textcolor{blue}{
The remaining sections of this paper are organized as follows.
Section \ref{sec:relatedWork} highlights the related work.
Section \ref{sec:systemintro} gives a brief overview of the water treatment control systems.
Section \ref{sec:adm} presents our proposed ML framework for anomaly detection in the SWaT system. Then, we apply our framework to a real-world SWaT system. 
We evaluate the performance of ML models and showed the results
in Section \ref{sec:exp}.
Finally, Section \ref{sec:5} concludes the paper and offers insights for future research.
}

\section{Comparison with Related Work}\label{sec:relatedWork}

The majority of research on adversarial machine learning has focused on the computer vision area. However, we believe it is important to extend prior work to other domains, including cyber-physical systems, which are vulnerable to real-world attacks \cite{rosenberg2020adversarial}. A study reported in \cite{umer2021attack} \textcolor{red}{used machine learning to generate} attack patterns for an operational water treatment plant. They used the same SWaT attack dataset that we have used in our proposed study. They employed Association Rule Mining (ARM) to generate a large number of attack patterns for SWaT. These attack patterns can later be used as a dictionary for signature-based anomaly detection.  However, our proposed study creates perturbations in the attack data. The perturbed attack data is used to create more robust supervised machine learning models for anomaly detection. The literature related to attacker models for ICS and model-based tools for SWaT risk assessment is described in the subsequent subsection.

\subsection{Attacker Models for ICS}

According to \cite{carlini2017towards}, adversarial samples were generated using three distance metrics: ${L_0}$, ${L_2}$, and $L_\infty$. These samples were created using \textcolor{blue}{limited-memory Broyden-Fletcher-Goldfarb-Shanno (L-BFGS), fast gradient sign method (FGSM), and JSMA, utilizing the modified National Institute of Standards and Technology (MNIST) database and Canadian Institute for Advanced Research (CIFAR) datasets}. In a recent study, the authors generated adversarial examples using an upgraded projected gradient descent (PGD) and an upgraded Carlini and Wagner (C\&W) method. The authors claim that both proposed algorithms required less time to generate adversarial examples \cite{huang2020two}.
 

\subsection{Model-based Tools for SWaT Risk Assessment}

Besides the computer vision domain, various techniques have been used for generating adversarial samples in the cybersecurity domain. In \cite{pacheco2021adversarial}, the authors generated adversarial samples using popular adversarial deep learning attack methods, such as JSMA, FGSM, and Carlini Wagner (CW), with modern IDS datasets (UNSW-NB15 and Bot-IoT). The study reported in \cite{abadi2021adversarial} generated adversarial samples using FGSM against condition-based maintenance (CBM) capabilities, evaluating the performance of a CBM system under attack. 

When attacking an autoencoder IDS, the authors in \cite{erba2019real} generated adversarial ICS attacks by substituting original data with readings within the normal sensor range. Although each sensor reading could be replaced from an arbitrary initial value to a value within the normal range, the perturbation applied in this method could be significantly large. Using the SWaT dataset, the authors in \cite{kravchik2020poisoning_BGU_AsafShabtai} introduced stealthy poisoning during the training phase to avoid detection in the test phase. They developed attacks for a residual signal threshold-based detector using seven attacks from the dataset.

A new adversarial attack method, called the Selective and Iterative Gradient Sign Method, was proposed in \cite{gomez2021crafting}, which required less time compared to the basic iterative method (BIM). The authors in \cite{inoue2017anomaly}, \cite{kravchik2018detecting}, and \cite{goh2017anomaly} evaluated long short-term memory (LSTM) networks for detecting cyber-physical attacks in the SWaT infrastructure, achieving an optimal LSTM with an F1 score of 0.80.


\begin{figure}[b] 
    \centering
    \includegraphics[width=0.45\textwidth]{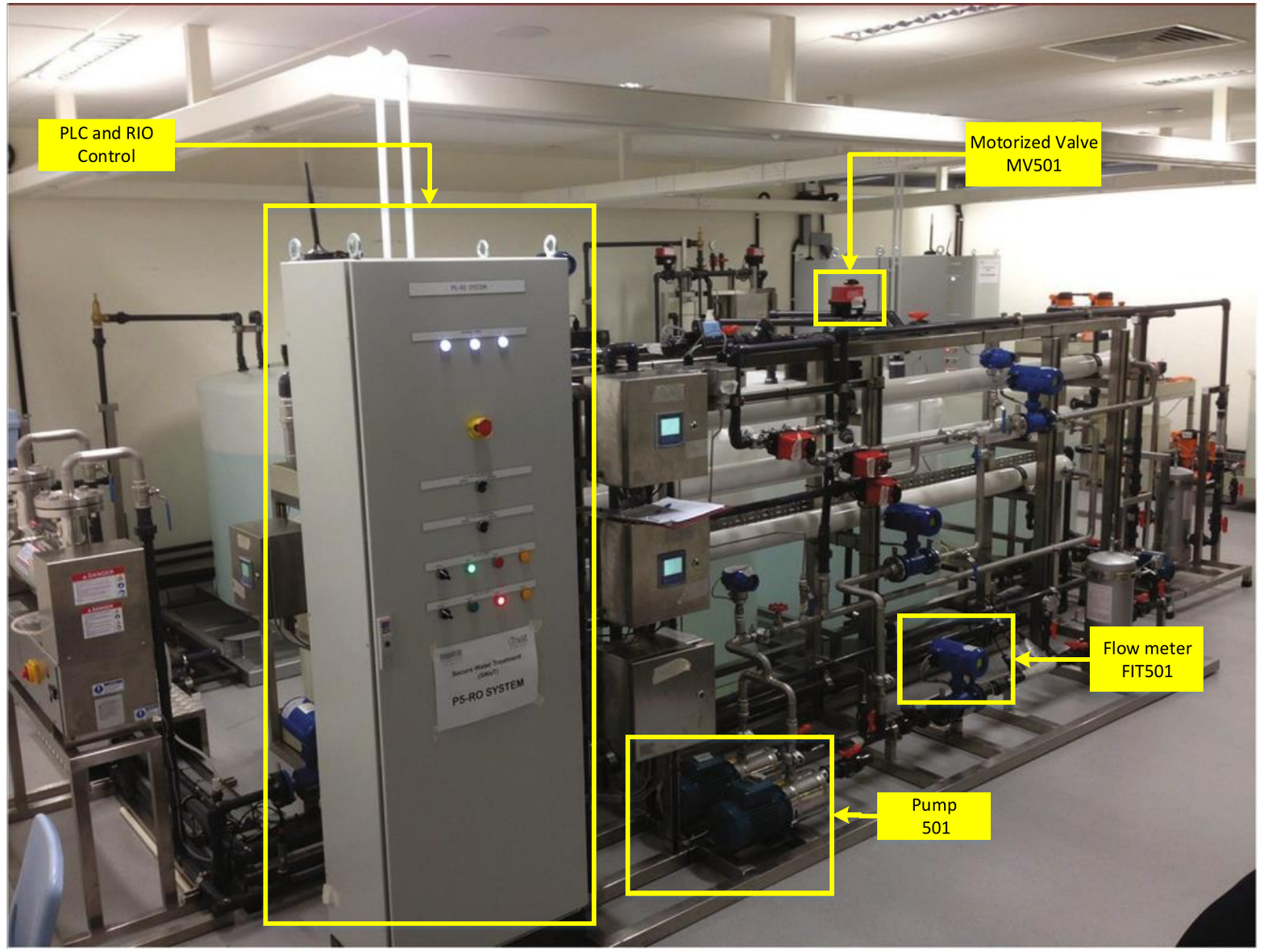}
    \caption{An overview of real-time SWaT system.}
    \vspace{-0.1in}
    \label{fig1:swatpic}
\end{figure}
\begin{figure*}[t] 
    \centering
    \includegraphics[width=0.75\textwidth]{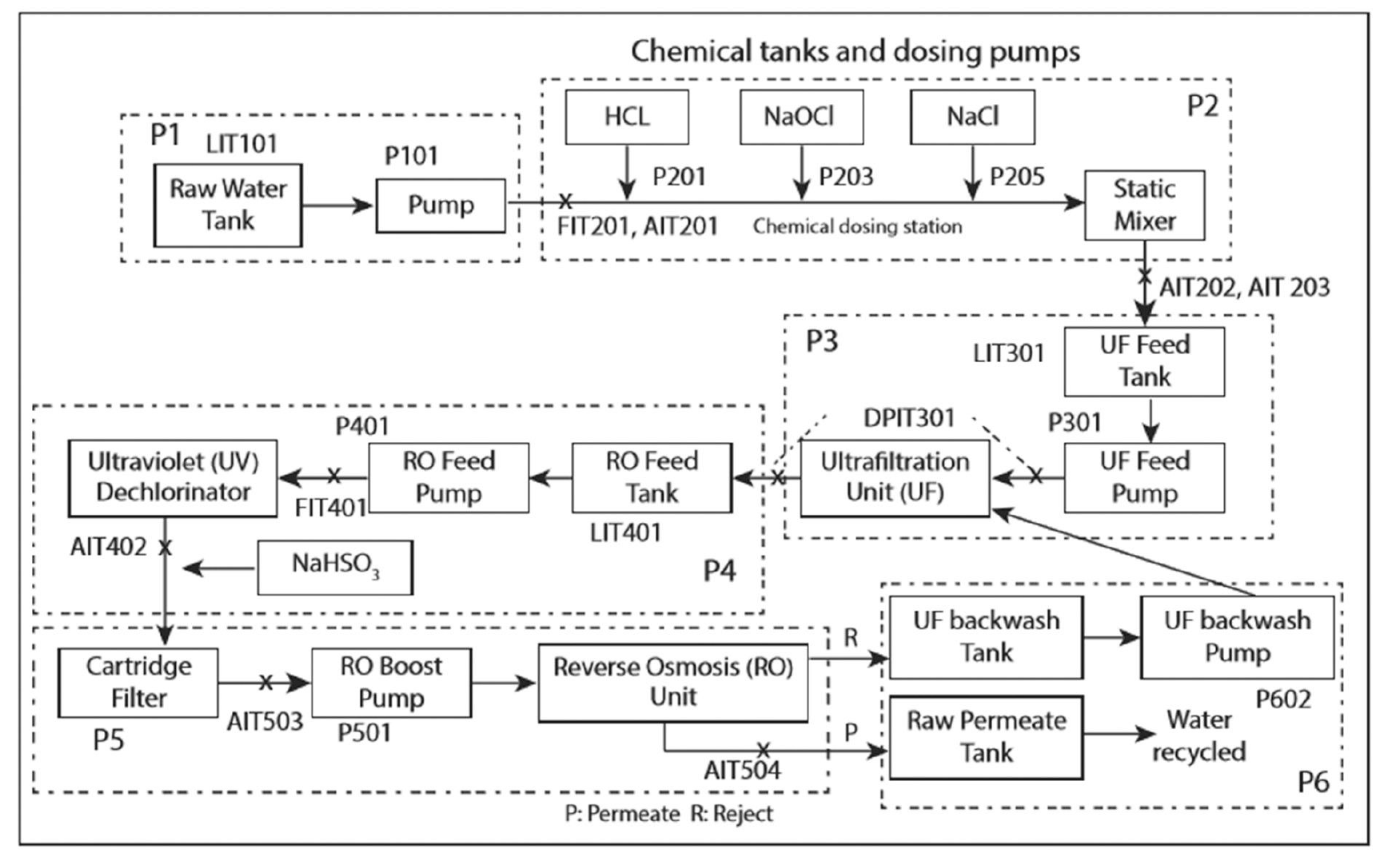}
    \caption{Illustration of SWaT testbed \protect\cite{goh2016dataset}.}
    \vspace{-0.1in}
    \label{fig2:swatprocess}
\end{figure*}

\section{SWaT Vulnerabilities and Attack Data}\label{sec:systemintro}

\subsection{SWaT Testbed}

The SWaT plant is an operational testbed available at the Singapore University of Technology and Design (SUTD) \cite{mathur2016swat}. A pictorial view of SWaT is shown in Figure \ref{fig1:swatpic}. It can produce five gallons of treated water per minute.  SWaT is a distributed control system (DCS) that consists of six stages, as shown in Figure \ref{fig2:swatprocess}. Each stage comprises a group of sensors and actuators, totaling 51 sensors and actuators. The sensors measure physical properties such as the water flow rate in pipes, the water level in tanks, and pressure.  Moreover, chemical monitoring sensors calculate a range of properties, including water conductivity, oxidation-reduction potential, and pH level. The actuators regulate the flow rate of water and chemical dosing. \textcolor{blue}{The six stages of SWaT, from P1 to P6, are summarized in Figure~\ref{fig2:swatprocess}.}

\begin{itemize}
    \item \textbf{P1}: ensures that the raw water tank has adequate water to supply the other processes. 
    \item \textbf{P2}: responsible for guaranteeing the quality standards of water. 
    \item The water is then sent to \textbf{P3}, once it has reached the required purity. In this stage, an ultra-filtration (UF) system with a fine filtration membrane removes any leftover unwanted items in the water. 
    \item All leftover chlorine is removed through dechlorination using ultraviolet (UV) rays in \textbf{P4}. The next step is to minimize the number of inorganic contaminants in the water.
    \item  The water in \textbf{P4} is then pumped into \textbf{P5} for reverse osmosis (RO).
    \item  The treated water is then distributed in \textbf{P6}.
\end{itemize}

\subsection{SWaT Attack Data}

A large number of researchers have used the SWaT testbed to examine cyber-attacks and their defense mechanisms for ICS~\cite{junejo2016behaviour, UMER_Azmi_2020, nedeljkovic2020detection, garg2021evaluation, ahmed2021machine, umer2021attack}. The SWaT dataset \cite{swatDataset} was generated by running the plant continuously for eleven days. During the first seven days, the plant was operated in a normal state. In the remaining four days, a total of 36 different attacks were performed on the SWaT testbed. The duration and goals of these attacks varied, with several attempting to cause underflow/overflow conditions in water treatment tanks, while others aimed to break pipes and stop filtration processes. The attacked points, according to the type of attack and stage, are presented in \cite{li2018anomaly}. One such attack targeted the level sensor LIT-101 of Stage P1, where the goal was to overflow the tank by manipulating the values of the LIT-101 sensor and turning off the pump P-101~\cite{sridhar_dataset_paper}.

\begin{figure}[!b]
    \centering
    \subfloat[]{\includegraphics[width=0.46\textwidth]{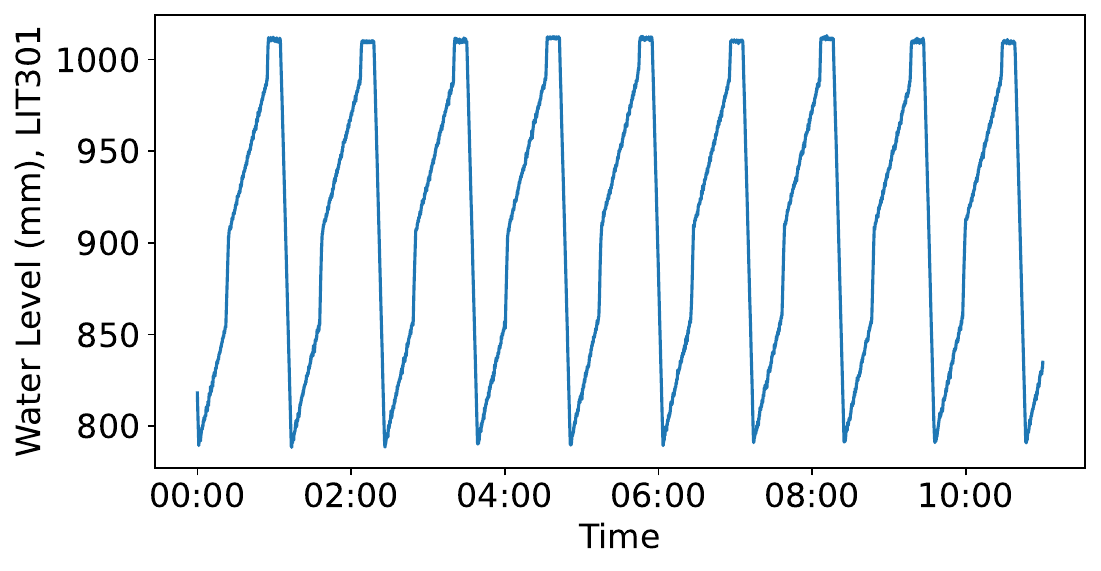}\label{fig:swatpic1}}%
    \hspace{-0.1\textwidth} 
    \subfloat[]{\includegraphics[width=0.46\textwidth]{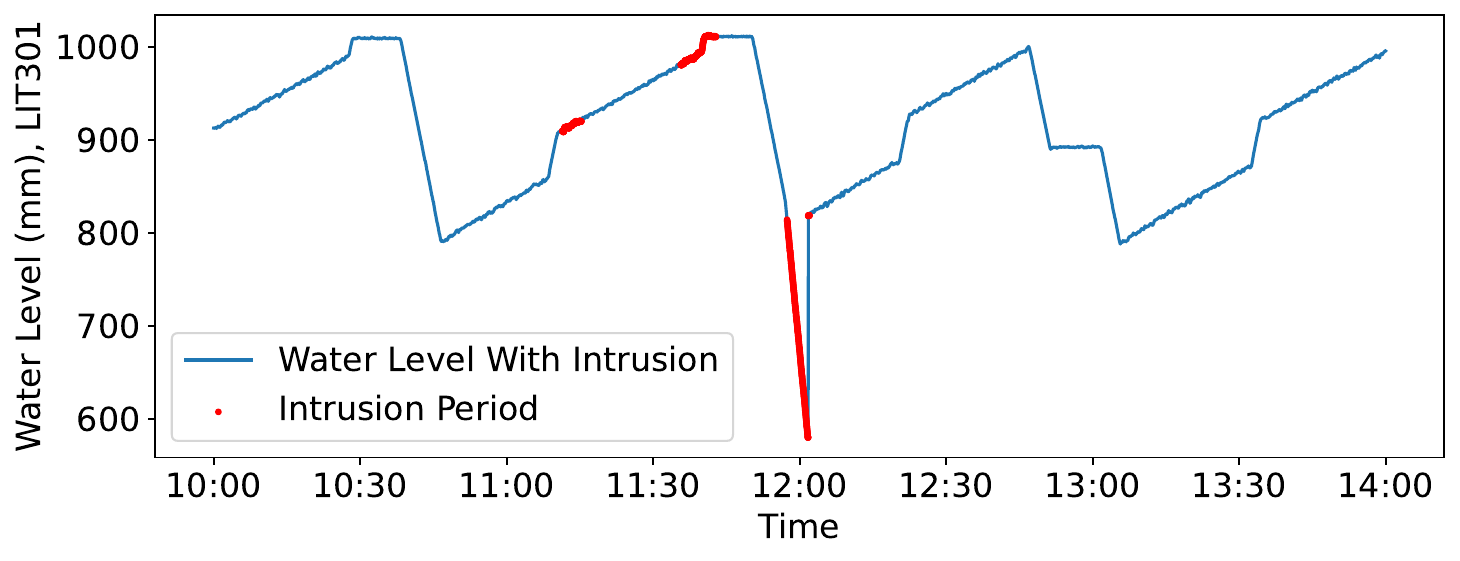}\label{fig:swatpic2}}%
    \caption{Example of normal (a) and attack (b) data, tank water level readings from sensor LIT301.}
    \label{fig:swatpic}
\end{figure}



\begin{figure}[t] 
    \centering
    \includegraphics[width=0.45\textwidth]{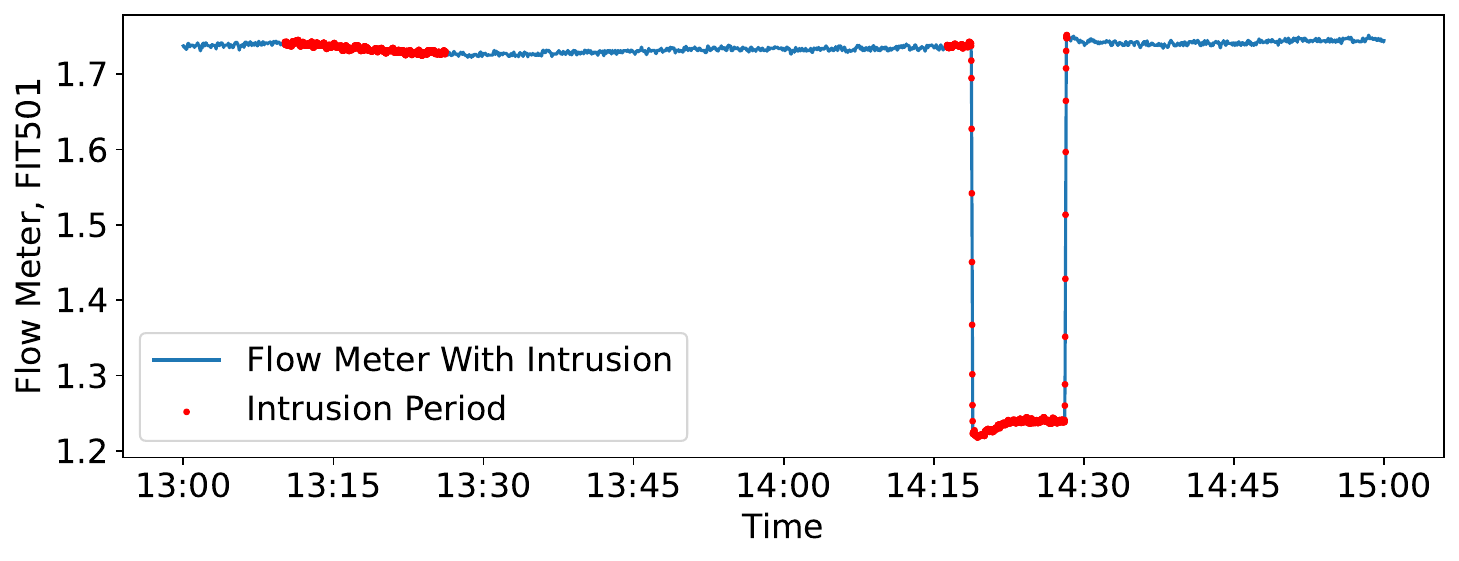}
    \caption{An example of attack data, showing flow meter readings from sensor FIT501.}
    \vspace{-0.1in}
    \label{fig4:swatpic}
\end{figure}

For instance, in Fig. \ref{fig:swatpic1}, the water level readings from sensor LIT301 clearly show the water consumption pattern from the tank. It illustrates the daily water consumption, where normal pumping events occur to fill the tank. The threshold must be maintained within the specified range. However, due to the cyber attack, the water level fell below the lower limit, posing a critical risk that requires immediate mitigation, as illustrated in Fig. \ref{fig:swatpic2}.

By analyzing the water level data, if an attack lowers the readings, it could falsely trigger (1) the pump to operate and (2) increase the risk of overflow. This inevitably results in more pumping events, leading to increased energy consumption. It is essential to monitor the float level and identify any false positives or false negatives. Similarly, in the case of the flow meter in Fig. \ref{fig4:swatpic}, the attack data reveals significant variations in flow, which can lead to potential malfunctions. In the following section, we propose an ML framework that uses attack data while incorporating adversarial sampling.

\subsection{Threat Model for Water Treatment Plant}

In this section, we formalize the types of attacks launched on our secure water treatment testbed (SWaT) as explained above. Essentially, the attacker's model encompasses the attacker's intentions and capabilities. The attacker may choose its goals from a set of intentions \cite{sridhar_compsac_2016}, including performance degradation, disturbing a physical property of the system, or damaging a component. These goals include under-flowing and over-flowing the water tank, bursting pipes, intentionally wasting water by passing it to the drain, and unnecessarily reducing the water in the tank.

\noindent It is assumed that the attacker knows the system dynamics and the control inputs and outputs. We consider a strong adversary who is able to launch both cyber and physical attacks. In an ICS, sensors, actuators, and PLCs communicate with each other via communication networks. An attacker can compromise these communication links in a classic \emph{Man-in-The-Middle~(MiTM)} attack~\cite{urbina_CCS2016limiting,SridharAdepu_AsiaCCS2016_L1Attacks,SAmin_2013_StealthyAtt_canal}, for example, by breaking into the link between sensors and PLCs. Besides false data injection in sensor readings via the cyber domain, an adversary can also physically tamper with a sensor to drive an ICS into an unstable state. Sensors can be connected to remote \emph{input/output} units via wired and wireless connections. A cyber attacker can remotely spoof sensor readings without needing physical access.

\noindent \textit{Data Injection Attacks:} For data injection attacks, it is considered that an attacker injects or modifies the real sensor measurements. The attacker's goal is to deceive the control system by sending incorrect sensor measurements. In this scenario, the level sensor measurements are increased while the actual tank level remains unchanged. This makes the controller think that the attacked values are true sensor readings, causing the water pump to keep working until the tank is empty, which can lead to the pump burning out. The attack vector can be defined as,

\begin{equation}\label{biasInjection_attack}
\bar{y}_k  = y_k  +  {\delta_k}.
\end{equation}

\noindent Where $y_k$ is the sensor measurement, $\bar{y}_k$ is the sensor measurement with the attacked value, and ${\delta_k}$ is the bias injected by the attacker. We can obtain a similar expression for an actuator attack as well.

\section{Application of Adversarial Machine Learning using JSMA}
\label{sec:adm}

Machine learning (ML) models are subject to adversarial attacks, where the attacker modifies input data to cause misclassification. An adversarial sample is carefully designed to disrupt the performance of a machine learning classifier. The attacker creates malicious inputs to fool the machine learning algorithms during the test phase~\cite{papernot2016limitations}. This is \textcolor{red}{one of the techniques in Adversarial Machine Learning (AML)}. Designing and developing robust ML-based algorithms to resist cyber-attacks is also part of this technique~\cite{cardenas2011}.

Specific characteristics of the attack model, the adversary, and the defenses are described in relevant research on AML. According to~\cite{barreno2006can}, such an attack has three primary characteristics. The attacker's capability is referred to as influence, which might be causative or exploratory, i.e., changing the input training data and learning classifier decisions after sending instances to the classifier. Security violation is the second property, which includes integrity, availability, and privacy. The third property is the attack's specificity: indiscriminate (the goal is to fail the classifier across a wide range of classes) and targeted (the goal is to fail the classifier for a specific instance). There are two types of potential attacks described by the threat models: black-box attacks, where the attacker is unaware of the model, and white-box attacks, where the attacker has knowledge of the model.

Adversarial samples can be generated using various approaches. The complexity, speed of generation, and performance of these methods vary. Manual perturbation of input data points is a naive method of creating such samples. However, manual perturbations are slower to develop and analyze than automated techniques. The Jacobian-based Saliency Map Attack (JSMA) was introduced by Papernot et al. in~\cite{papernot2016limitations}. The authors used JSMA for image recognition tasks. In the current study, we have used it to generate adversarial samples for time series data composed of different sensors and actuators of the SWaT.

The JSMA approach uses saliency maps to generate perturbations. The saliency map identifies the important features of input data for classification; if these features are changed, the target values will most likely be classified differently. A percentage of $(\theta)$ is perturbed using $(\gamma)$, i.e., the noise. The model then determines whether the introduced noise has led to misclassification by the targeted model. If the model's performance is unaffected by the noise, a new collection of features is chosen, and a new cycle begins until a saliency map that can generate the adversarial samples emerges. The technique acquires the Jacobian matrix as described in equation \ref{eq:jacobMatrix}, where $i$ is the input component and $j$ is the class derivative for input sample $x$.

\begin{equation}
\label{eq:jacobMatrix}
J_F(x) = \frac{\partial F(x)}{\partial x} = \left[ \frac{\partial F_j(x)}{\partial x_i} \right]_{i \times j}
\end{equation}

\begin{align}
\label{eq:saliencyMap}
S(x, t)_i = 
\begin{cases}
0, & \text{if } \frac{\partial F_t(x)}{\partial x_i} < 0 \\
& \text{or} \sum_{k \neq t} \frac{\partial F_k(x)}{\partial x_i} > 0 \\
\frac{\partial F_t(x)}{\partial x_i} \cdot \left| \sum_{k \neq t} \frac{\partial F_k(x)}{\partial x_i} \right|, & \text{otherwise}
\end{cases}
\end{align}

\begin{equation}
\label{eq:perturb}
x_i^{\text{new}} = x_i + \epsilon
\end{equation}

Here, $x_i$ represents the i-th feature of the input sample $x$.
In eq \ref{eq:saliencyMap} the saliency map is calculated, the input is iteratively modified by selecting the feature $x_i$ with the highest saliency score as described in eq \ref{eq:perturb}. 
Here, $\epsilon$ is the perturbation step size that is chosen to be small enough to maintain a gradual change in the perturbation.

In \textcolor{red}{Equations (\ref{eq:jacobMatrix}) and (\ref{eq:saliencyMap})}, \textbf{F} represents the output of the penultimate layer (related to the output of the softmax layer)~\cite{yuan2019adversarial}. The perturbation is selected, and the method is iterated until the target is misclassified or the maximum number of perturbed features is reached. If this fails, the algorithm proceeds to the next feature, which is then added to the perturbed sample. For instance, the authors in \cite{yuan2019adversarial} achieved a $97\%$ adversarial success rate while modifying only $4.02\%$ of the input features per sample. This procedure requires complete knowledge of the design and parameters of the target model~\cite{papernot2016limitations}.

\begin{figure*}[t] 
    \centering
    \includegraphics[width=0.8\textwidth]{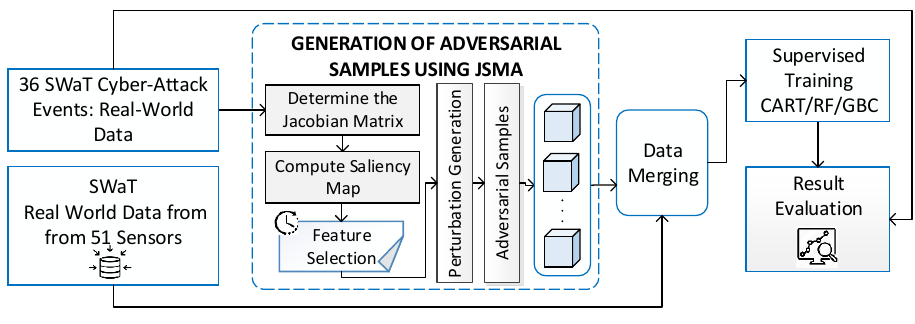}
    \caption{\textcolor{red}{Proposed workflow for evaluating 36 SWaT cyber-attack events using real-world data from 51 sensors. The process includes supervised training with CART/RF/GBC and adversarial sample generation via JSMA.}}
    \vspace{-0.1in}
    \label{fig4:overflow}
\end{figure*}



 
\section{EXPERIMENTAL SETUP} \label{sec:exp}

We have used CleverHans V.3.0.0 library 
to generate the adversarial samples more specifically to implement JSMA~\cite{papernot2016cleverhans}. The Tensorflow V.1.14.0 \cite{tensor45166}, and Keras V.2.0.0 \cite{chollet2015} were used for the pre-processing, experimental evaluations, and analysis.  Our study is based on a binary classification problem as the SWaT dataset comprises two classes i.e., attack or normal. Based on this we trained multiple models using various algorithms. Here we have summarized the experimental evaluation for the top three algorithms i.e., Classification And Regression Trees (CART), Random Forest (RF), and Gradient Boosting Classifier (GBC).

\subsection{Data Pre-processing}

It is crucial to structure the dataset in the pre-processing step, especially for supervised machine learning. To transform nominal values into numerical values, we used label encoding. For instance, in the SWaT dataset \cite{swatDataset}, the target label has two nominal values, namely attack and normal, which need to be mapped into their respective numerical values, with normal and attack mapped to 0 and 1, respectively. Since the dataset contains features with different distributions, min-max normalization was applied to all features after label encoding. For min-max normalization, 0 and 1 were chosen as the minimum and maximum range, respectively.

\subsection{Adversarial Sample Generation}
We used two publicly available SWaT datasets. There are 51 attributes in the SWaT datasets. Among these attributes, 25 are related to sensor readings and the remaining 26 are related to actuator readings. The first SWaT dataset was collected during the normal operation of the plant, which we refer to as the normal dataset. This dataset contains 410,400 transactions and was collected at a frequency of one transaction per second. The second dataset was collected by performing 36 attacks at different time instances. We refer to this dataset as the attack dataset. This dataset contains 449,919 transactions and was also collected at a frequency of one transaction per second. The attack dataset contains 53,900 anomalous and 396,019 normal transactions. We used the attack dataset to generate the adversarial samples using JSMA. The Cleverhans library was used for the implementation of JSMA. A Multi-Layer Perceptron (MLP) was chosen as the pre-trained underlying model for the generation of adversarial samples. We generated 112,480 adversarial samples using the proposed approach.

\subsection{Supervised Model Training using Adversarial Samples}
The 112,480 adversarial samples generated earlier were merged with the SWaT normal dataset, which contains 410,400 transactions. Therefore, the merged dataset contains 522,880 transactions. This merged dataset was used to train the supervised models. In particular, we used CART, RF, and GBC for this purpose. Among these algorithms, CART is a simple decision tree algorithm, while RF and GBC are ensemble algorithms that use decision trees. The ensemble algorithms build a collection of classifiers and take a vote from each classifier's predictions to classify new data points \cite{Dietterich00ensemblemethods}. The main purpose of training these models is to test the effectiveness of generative adversarial samples on attack detection. Therefore, we evaluated the performance of the trained models on the SWaT attack dataset. The complete process of adversarial sample generation, model training, and evaluation is described in Figure \ref{fig4:overflow}.




\subsection{Results}

\begin{table*}
\caption{Results from various modern ML classifiers.}
\label{table1}
\centering
{
\begin{tabular}{|c|c|c|c|c|c|c|c|c|c|} 
\hline
\multirow{2}{*}{\textbf{Classifier}} & \multicolumn{3}{c|}{\textbf{Accuracy}} & \multicolumn{3}{c|}{\textbf{Precision}} & \multicolumn{3}{c|}{\textbf{Recall}} \\ 
\cline{2-10}
& \textbf{Worst} & \textbf{Average} & \textbf{Best} & \textbf{Worst} & \textbf{Average} & \textbf{Best} & \textbf{Worst} & \textbf{Average} & \textbf{Best} \\ 
\hline
\textbf{CART} & 0.88 & 0.90 & 0.95 & 0.88 & 0.90 & 0.94 & 0.99 & 0.99 & 0.99 \\ 
\hline
\textbf{RF} & 0.88 & 0.88 & 0.88 & 0.88 & 0.88 & 0.88 & 1.0 & 1.0 & 1.0 \\ 
\hline
\textbf{GBC} & 0.95 & 0.95 & 0.95 & 0.95 & 0.95 & 0.95 & 1.0 & 1.0 & 1.0 \\ 
\hline
\end{tabular}
}
\end{table*}

\begin{table*}
\caption{FPR and F1-Score for various modern ML classifiers.}
\label{table2}
\centering
{
\begin{tabular}{|c|c|c|c|c|c|c|} 
\hline
\multirow{2}{*}{\textbf{Classifier}} & \multicolumn{3}{c|}{\textbf{FPR}} & \multicolumn{3}{c|}{\textbf{F1-Score}} \\ 
\cline{2-7}
& \textbf{Worst} & \textbf{Average} & \textbf{Best} & \textbf{Worst} & \textbf{Average} & \textbf{Best} \\ 
\hline
\textbf{CART} & 0.99 & 0.80 & 0.41 & 0.94 & 0.95 & 0.97 \\ 
\hline
\textbf{RF} & 1.0 & 1.0 & 1.0 & 0.94 & 0.94 & 0.94 \\ 
\hline
\textbf{GBC} & 0.37 & 0.37 & 0.37 & 0.97 & 0.97 & 0.97 \\ 
\hline
\end{tabular}
}
\end{table*}


\begin{table*}
\caption{Confusion Matrix of CART, RF, GBC}
\label{table3}
\centering
\begin{tabular}{|c|c|c|c|c|c|c|c|c|c|c|} 
\hline
& \multicolumn{3}{|c|}{\textbf{Worst Score}} & \multicolumn{3}{c|}{\textbf{Average Score}} & \multicolumn{3}{c|}{\textbf{Best Score}} \\ 
\hline
& \begin{tabular}[c]{@{}c@{}}\textbf{~Predicted Class~→}\\\textbf{True Class~↓}\end{tabular} & \textbf{Normal} & \textbf{Attack} & \begin{tabular}[c]{@{}c@{}}\textbf{Predicted Class~\textbf{→}}\\\textbf{~True Class~\textbf{↓}}~\end{tabular} & \textbf{Normal} & \textbf{Attack} & \begin{tabular}[c]{@{}c@{}}\textbf{Predicted Class~}\textbf{→}\\\textbf{~True Class~\textbf{↓}}\end{tabular} & \textbf{Normal} & \textbf{Attack} \\ 
\hline
\multirow{2}{*}{CART} & \textbf{Normal} & 395935 & 84 & \textbf{Normal} & 395935 & 84 & \textbf{Normal} & 395954 & 65 \\ 
\cline{2-10}
& \textbf{Attack} & 53854 & 46 & \textbf{Attack} & 53854 & 46 & \textbf{Attack} & 21927 & 31973 \\ 
\hline
\multirow{2}{*}{RF} & \textbf{Normal} & 396019 & 0 & \textbf{Normal} & 396019 & 0 & \textbf{Normal} & 396019 & 0 \\ 
\cline{2-10}
& \textbf{Attack} & 53900 & 0 & \textbf{Attack} & 53900 & 0 & \textbf{Attack} & 53900 & 0 \\ 
\hline
\multirow{2}{*}{GBC} & \textbf{Normal} & 396019 & 0 & \textbf{Normal} & 396019 & 0 & \textbf{Normal} & 396019 & 0 \\ 
\cline{2-10}
& \textbf{Attack} & 20236 & 33664 & \textbf{Attack} & 20178 & 33722 & \textbf{Attack} & 19840 & 34060 \\ 
\hline
\end{tabular}
\end{table*}

The performance of the previously trained models was tested using the SWaT attack dataset. The results are shown in Tables \ref{table1}, \ref{table2}, and \ref{table3}. The SWaT attack dataset is highly imbalanced, with normal-class samples far outnumbering attack-class samples. Therefore, accuracy alone can be misleading. Instead, we evaluated the proposed technique using additional metrics such as precision, recall, false positive rate (FPR), and F1-Score. All these metrics are defined in the following.

	
\noindent Where True Positive (TP) represents the attack instances that are correctly classified as attack. The True Negative (TN) represents the normal instances that are correctly classified as normal. The False Positive (FP) represents the normal instances that are incorrectly classified as attack. The False Negative (FN) represents the attack instances that are incorrectly classified as normal. The TP, FP, TN, and FN form the confusion matrix of the ML classifier. \textcolor{blue}{
Note that in calculating the F1 score and accuracy, we determine the true positive (TP), false positive (FP), true negative (TN), and false negative (FN) at a trace level:
\begin{itemize}
\item{\textbf{True Positive: }} A faulty trace that is flagged instances that are correctly classified as normal.
\item{\textbf{False Positive: }} Either a normal trace that is flagged as faulty or a faulty trace that is flagged as faulty before the time of occurrence of the fault.
\item{\textbf{True Negative: }} represents the attacked instances that are correctly classified as attacked.
\item{\textbf{False Negative: }} A faulty trace that is not flagged as faulty.
\end{itemize}
}


We have presented three scenarios: worst, average, and best case for accuracy, precision, recall, FPR, and F1-Score for each classifier, as shown in Tables \ref{table1} and \ref{table2}. Accuracy represents the overall performance of the classifier. CART and GBC achieved a maximum accuracy of 95\%. However, the average scores of both classifiers differ, with CART at 90\% and GBC at 95\%. As mentioned earlier, the current problem is class imbalance; therefore, accuracy alone is not sufficient to assess the performance of classifiers. We also calculated the precision and recall of all the classifiers. Precision here represents the performance of classifiers in identifying the normal instances in the dataset, while recall represents the identification of normal instances with respect to the total normal instances in the dataset. There is a trade-off between precision and recall. For this purpose, we use another metric, the F1-score, which is the harmonic mean of precision and recall. Improving the F1-score helps maintain the balance between precision and recall. Additionally, we calculated the false positive rate for each classifier, as a high rate of false positives makes the IDS impractical for real-world applications. For an in-depth evaluation of the classifiers' performance, the confusion matrices of each classifier are given in Table \ref{table3}.

From the confusion matrix of RF in Tables \ref{table1} and \ref{table2}, it is evident that the classifier was unable to differentiate between attack and normal instances. Consequently, it classified all attack instances as normal, even though its accuracy is 88\%. The confusion matrix of CART in Table \ref{table3} shows that its performance was better than RF in detecting attack instances. The confusion matrix of GBC in Table \ref{table3} shows that it performed better than CART not only in detecting attacks but also in classifying normal instances. These results highlight that the examples generated by JSMA proved useful in improving the accuracy of detectors without needing to be trained on attack data.

\section{Conclusions}\label{sec:5}


\textcolor{red}{In this paper, we assessed the quality of the malicious data created by the JSMA attack method, using the SWaT dataset as a testbed.}
Although JSMA was originally designed to create perturbations for image data, it was successfully exploited for time series data. 
\textcolor{red}{Machine learning classifiers often lack sufficient data to defend against attacks.}
\textcolor{red}{Our results show that the proposed approach improves the performance of these classifiers against previously unseen attacks.}
Future work will focus on enhancing the robustness of IDS against a wider range of adversarial attacks. This includes exploring other adversarial attack methods and developing more sophisticated defense mechanisms. Additionally, we plan to extend our evaluation to other ICS datasets to further validate the effectiveness of our approach. Investigating the integration of ML-based IDS with other security measures in ICS will also be a key area of future research.

\section*{Acknowledgement}
This research is supported in part by the National Research Foundation, Singapore, under its National Satellite of Excellence Programme “Design Science and Technology for Secure Critical Infrastructure: Phase II” (Award No: NRF-NCR25-NSOE05-0001). Any opinions, findings and conclusions or recommendations expressed in this material are those of the author(s) and do not reflect the views of National Research Foundation, Singapore.

\bibliographystyle{unsrt}
\bibliography{references.bib}
\end{document}